\documentclass[conference]{IEEEtran}
\ifCLASSINFOpdf
\else
\fi

\usepackage{url}

\usepackage{booktabs} 
\usepackage{setspace}

\usepackage{xspace}
\newcommand{\cf}{{ComFlux}\xspace}

\usepackage{caption}
\usepackage{subcaption}
\usepackage{epstopdf}

\usepackage{graphicx}
\graphicspath{{figures/}{charts/}}

\usepackage{array}

\usepackage{color}

\begin{document}
%

\title{ ComFlux: External Composition and Adaptation of Pervasive Applications}

\author{
\IEEEauthorblockN{Raluca Diaconu\IEEEauthorrefmark{1}, Jie Deng\IEEEauthorrefmark{2}, Jean Bacon\IEEEauthorrefmark{1}, Jatinder Singh\IEEEauthorrefmark{1}}\vspace{-.5cm}\and\and
\IEEEauthorblockA{\IEEEauthorrefmark{1}University of Cambridge, UK\\
Email: \{raluca.diaconu, jean.bacon, jatinder.singh\}@cl.cam.ac.uk}
\and
\IEEEauthorblockA{\IEEEauthorrefmark{2}Queen Mary University of London, UK\\
Email: j.deng@qmul.ac.uk}
}


%


\maketitle


\begin{abstract}


Technology is becoming increasingly pervasive. At present, the system components working together to provide functionality, be they purely software or with a physical element,
tend to operate within silos, bound to a particular application or usage.
This is counter to the wider vision of pervasive computing, where a potentially limitless number of applications can be realised through the dynamic and seamless interactions of system components. We believe this application composition should be externally controlled, driven by policy and subject to access control. We present ComFlux, our open source middleware, and show through a number of designs and implementations, how it supports this functionality with acceptable overhead.
\end{abstract}

%
\IEEEpeerreviewmaketitle

\section{Introduction}
\label{sec:intro}
This century has witnessed increasingly specialised technology
becoming better at dealing with specific tasks such as sensing, automation, and mobility.
Sensors, actuators, mobile devices and wearables, integrated and driven by a range of software services, increasingly become incorporated into our everyday activities, and commercial promotion of the Internet of Things (IoT) will increase these tendencies. 

The goal of pervasive computing~\cite{percomvision, CaceresF12} is transparent integration of technology components into the environment to create contextual functionality ---``a cyber-physical continuum''. Seamlessness is key in users' interacting with pervasive technology; 
\emph{components} need to interact smoothly with users, with each other and with the environment to achieve the functionality required.
This involves exchange of data, between system \emph{components}, be they physical (sensors, actuators, mobile devices) or software (apps, databases, web services). 

To date, functional requirements have typically been met by composing software and hardware components into ``silos of things''~\cite{silos}, managed in isolation.
We argue that such component functionality could potentially be used for many diverse purposes, outside their initial scope.
To achieve this, means are needed to `instruct' components how, when and with whom they should interact, taking into account usability, modularity and access control. This should happen dynamically at runtime, without requiring application-level intervention.
This would pave the way for new functionality, where existing components can be leveraged across application boundaries, in ways and for reasons not envisaged by their designers, and increase the longevity of components through updates and new uses. Section~\ref{sec:context} expands on these requirements and challenges.

To support this vision, we propose a system architecture that \emph{enables an external management regime}, whereby component interactions, communication methods and security constraints can be dynamically defined, extended and updated at runtime ---without requiring redeployment or changes to the application logic of the components themselves, and without imposing constraints on system design (Section~\ref{sec:cf}).
We show that such capabilities can be built into a messaging infrastructure, by presenting \emph{ComFlux}, 
an open-source, proof-of-concept middleware (Section~\ref{sec:cf}).
The aim of ComFlux is to support reconfiguration relating to the exchange of data in pervasive environments. We use ComFlux to demonstrate the potential of an external management regime in facilitating a range of dynamic, pervasive-computing scenarios, including those where control is user-(individual) and environment-centric. 

External control and communication management raise important security concerns, in that only authorised components should be able to perform reconfiguration actions. 
IoT and pervasive computing may deal with sensitive data and this may be subject to law if people are identified. Breaches in personal data exposure \cite{unicorns} and malicious behaviour \cite{ronen2016iot} are potentially harmful, unintended outcomes of IoT. Approaches to external reconfiguration should address these vulnerabilities, enforcing access control over the exchange of data (Section~\ref{sec:challenges}).

The components used for particular functionality may change over time as a result of updates or changes in context, e.g., as people move about, their surroundings evolve.
To support this, \cf has a modular design and provides mechanisms to extend and adapt the capabilities of an application at runtime including functional evolution of applications and software lifecycle updates.
Different modules can be plugged in and out to allow flexible functionality (Section~\ref{sec:modules}).

\section{Requirements and challenges}
\label{sec:context}
In this section we explore a wider vision of ubiquitous computing and highlight the challenges introduced by heterogeneous and dynamic environments.
We use smart cities and pervasive gaming to motivate our ``command and control'' approach to achieving novel functionality.

\subsection{Pervasive computing scenarios} 
We are moving towards an environment where virtual and physical are strongly intertwined, driven by advances in mobile technologies, cloud-based services and the emerging IoT. Pervasive technologies are having a great impact in areas ranging from consumer products to smart city projects, creating interest from users, industry and researchers.

Current cityscapes have increasingly many sensors e.g., temperature, presence, pollution, and more automated functionality gives the potential for innovative applications, provided citizens are well-informed and perceive benefit.
Digitally enhancing a space can help to make it more attractive. This involves adding more sensors, more actuators, and more ways for the user to interact and engage with it.

\emph{Pervasive games} might also extend the virtual gaming experience into the real world, and these are not confined to a space and time~\cite{Montola2009, BenfordML05}. 
Their deployment and gameplay can use the available and accessible infrastructure of a city including not only geographical, public transport, pollution and traffic information, but also information about energy and water distribution, development plans,
events and accidents~\cite{Schouten2017}.

Pervasive games can offer an interactive approach to understanding the history of a city~\cite{MixedHeritage} or to inform people's choices and responsibilities as consumers~\cite{RicciPTC15, Schouten2017}.
An example is \emph{The Water Must Flow}, a strategic game of resource management about using water in common and private areas of a neighborhood: it tasks players with managing plots of land in relation to rainwater, flooding, draught, and other similarly water-related~\cite{Schouten2017} issues.
Urban gaming could benefit greatly from seamless interaction via deployed infrastructure. 

In emergency situations, maximum functionality should be acquired from existing systems and commonly used components by automatically adapting their functionality and overriding access control to data as appropriate, provided such actions are audited.
We envisage that if a person falls ill, the surrounding devices should have their connections changed dynamically to inform medical staff, and then to allow rescue services to access data; for instance, a smart door lock may temporarily allow any doctor to pass through.


\subsection{The need for external command and control} 
\label{sec:contextexamples}
By their nature, pervasive computing systems are dynamic. Technology is not confined to a single place or context; instead, system components are mobile and the contexts under which they operate may vary.  
Given their complexity, pervasive systems pose the challenge of adapting to new environments and usage contexts. While the broader vision is of plug-and-play modules, dynamic reconfiguration and seamless access control, current solutions rely on pre-programmed functionality, where the only capacity for management is in terms of what applications expose through their in-built application interfaces (API), see Section~\ref{sec:rw}.

Pervasive systems are also heterogeneous and highly distributed. 
To accommodate this, off-the-shelf IoT products generally come with a complete solution, involving a connected device, a cloud service, and a suite of control and visualisation apps for web or smart-phone. Often they require an account registration and sometimes a periodic fee. Popular examples range from wearables (e.g., Fitbit~\cite{fitbit}) to IoT home solutions (Philips Hue~\cite{hue}, Nest~\cite{nest}, Apple HomeKit~\cite{homekit}, etc.) to platforms for specific (particularly industrial) application contexts (e.g., ThingWorx~\cite{thing}). 
Current approaches target a restricted system and/or application. In some cases, access to some sort of API may extend their usability, but achieving new functionality relies on  pre-programming the functionality that manages components' capabilities: with whom and how it interacts.  

To get the most out of existing devices and services it must be possible to reuse them in different circumstances, to adapt them to new purposes and to mediate the communication between them. 
We argue that the capabilities enabled by the API functionality need also to be available externally, at runtime, and show in Section~\ref{sec:cf} how to achieve this.
By enabling runtime reconfiguration, application development is simplified: application logic is decoupled from the reconfiguration and existing infrastructure can be reused, e.g., by interfacing with open services or public sensors. 

It is a challenge to manage and coordinate complex and large-scale systems, such as smart cities, as the answers to  \emph{how, when and why} things should interact, change over time.
Thus, applications need to be \emph{aware of the context} they operate in, that is, they need a mechanism that allows them to receive input and adapt to changes in the environment~\cite{PereraZCG14, MakrisSS13, YauK04}, as well as individual preferences which change over time. 
As systems become increasingly heterogeneous, dynamic and complex, the emerging range of applications cannot rely on pre-programmed functionality.  
Instead, new applications must be able to discover and interact within new contexts dynamically, at runtime, and be managed externally.  
The key aspect in enabling transparent integration with the environment is to leverage appropriate \emph{external command and control} over components' interactions, communication methods and security constraints. The functionality must be dynamic and the control and command interface must be available at runtime.

\subsection{External control: Requirements and challenges} 
\label{sec:challenges}
The real value of pervasive computing lies beyond its myriad technologies; it emerges from their operation within a context. The power of a technology component may be revealed when  \emph{reused} and \emph{adapted} to fit new contexts. 
Nowadays, driven by human dynamics, heterogeneous components and their mobility create increasingly complex environments in which dynamism and creativity can bring great value.
We now articulate pivotal challenges for future applications and propose ways to address them.



\paragraph{Capabilities management and communication coordination}
Pervasive technologies should offer control over their capabilities. They will allow external control, from users and other objects in the environment, over their functionality and their connections.
Consider a museum application that detects and coordinates interaction between people and its installations and services. On arrival, people will be able to discover the types of installations and load mechanisms to communicate with them. 
Inside, the environment detects people's presence, communicates with them to help coordinate groups, and may reconfigure its components on-the-fly to offer the participants contextual information. 

\paragraph{Seamless system composition} 
At urban scale, smaller-scale systems (and their constituent components) should be composable to bring new functionality. Resources should be used when and where needed, but no more, e.g., at night, lights could be activated only when people are detected, and lighting could ``follow them''.

\paragraph{Component reuse and repurposing}
Large-scale systems could be enabled by reusing and adapting smaller-scale components beyond their initial purpose.
Connected smart cities could bring awareness of their problems or draw attention to their landmarks, using real time sensor information and actuators. 
Individuals and groups could customise the use of the components in the environment to enhance particular activities, e.g., during a festival, as part of a game and so forth. 
Because of their scale and complexity such applications will have to discover and interact with components (devices and services), already part of the city's infrastructure. For example, when high pollution is detected, alerts and displays can indicate the danger zones, and routes for pedestrians and cyclists that avoid the pollution can be highlighted.

\paragraph{Long-lived components}
Updating the capabilities of potentially long-lived, deployed components is important in an IoT context, to add or change functionality as appropriate. 
Importantly, reuse and repurposing extends the life of the components themselves, as new uses are found. 

\paragraph{Security}
\label{sec:security}
External control over a component's connections and capabilities raises significant security concerns. 
Access control and authentication need to enforce that only authorised entities perform external reconfiguration actions. 

\paragraph{Automatic context adaptation}
The environment must sometimes be able to coordinate itself without human involvement. In an emergency, when human intervention may be too slow or not available, e.g., people are trapped under a collapsed building or are suffering cardiac arrest, the devices must collectively make decisions and reconfigure their connections and even their function; see further Section~\ref{sec:model}.
Emergency override of access control must always be audited.

In Section~\ref{sec:cf} we describe
\cf, an open source infrastructure (middleware)
to support externally-driven, dynamic reconfiguration under access control. 
In Section~\ref{sec:demo} we show how \cf meets the above challenges.


\section{ComFlux}
\label{sec:cf}

\subsection{The ComFlux model for external, dynamic reconfiguration}
\label{sec:model}

The \cf model allows components to be instructed externally and at runtime \emph{with whom} they should interact, \emph{what are the means} to communicate with new components, and \emph{when}, under which circumstances they should interact.

\begin{figure}[htbp]
   \begin{center}
   \includegraphics[width=.7\linewidth, trim={3cm 21.5cm 3cm 2.1cm},clip]{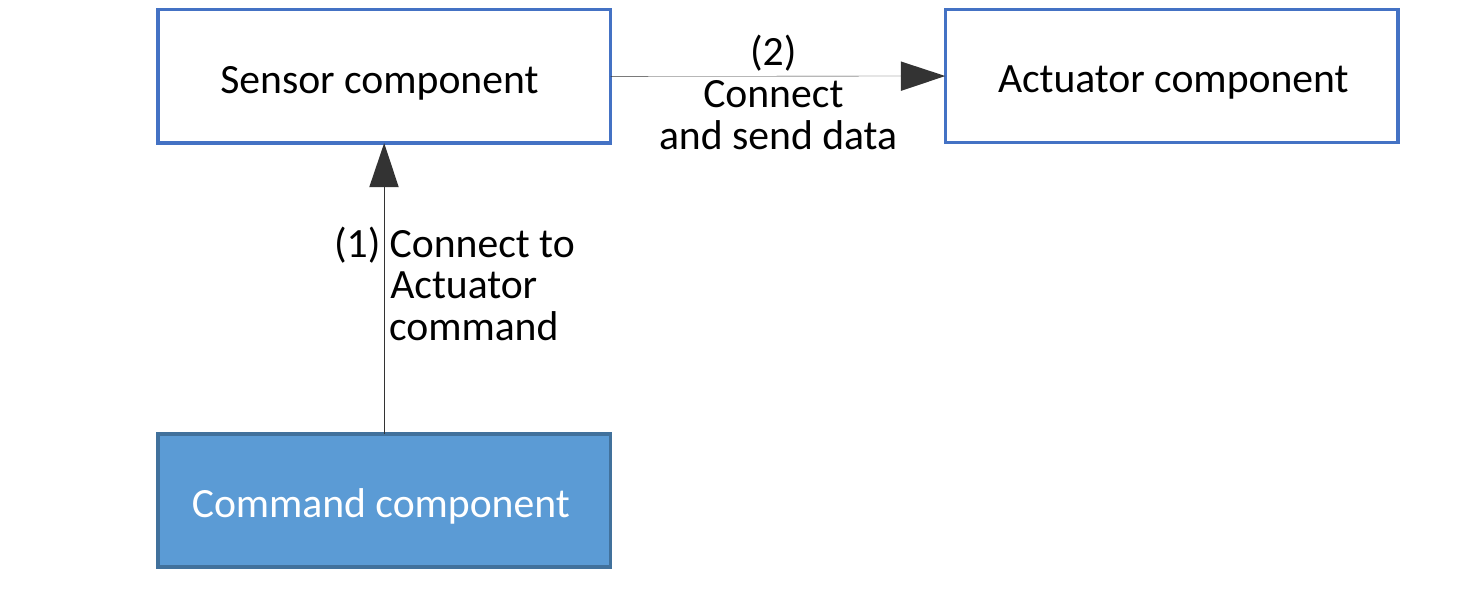}
   \caption{\small Example of external reconfiguration} 
   \vspace{-4mm}
   \label{fig:model}
   \end{center}
   \end{figure}

A ComFlux \emph{component} is a process, an application or a service. In a pervasive environment it can be the process running on an embedded device that interacts with other components by exchanging data, often wirelessly, or there might be a series of components, for instance, representing different applications on a mobile phone.
A pervasive component's functionality is defined by the data it exchanges. 
%
%
Components can reconfigure themselves, via their API, and also certain components can control others.  
Our model is based on the following key external runtime controls: 

1) \emph{External communication control}: an authorised component can instruct another component to connect or disconnect to\slash from some specified component(s) and transmit or receive data, see Fig.~\ref{fig:model} and Sections~\ref{sec:messaging}, \ref{sec:external}.


   2) \emph{External component (re)configuration}: an authorised component can reconfigure other components' functionality.
Fig.~\ref{fig:selectiona} shows reconfigurations to change the functionality of a specific application; see also Section~\ref{sec:modules}.
%

The mechanism for dynamic reconfiguration allows components to manage \emph{how} other components communicate and function. External reconfiguration capabilities can involve changing communication protocols in a plug-and-play manner and adding\slash removing access privileges over other components; e.g., a command component can instruct a component to load a new communication interface for UDP, instead of TCP connections, if its battery runs low, or instruct a sensor component to relax its data access permissions in an emergency.

External control might also include commands that specify \emph{what} data to pass and/or \emph{when} to interact, e.g., a component may add filters to its send\slash receive messages to conform to a pattern, such as only extreme temperatures. Temporal filters may specify when to receive messages, e.g., at what frequency. 

Developers of applications need not include and foresee interactions in all contexts; instead, the functionality is dynamic.
The ability to coordinate components is crucial, as is the ability to change and adapt their operation. By providing the means for runtime external configuration of key functionality --given permission-- our model provides greater flexibility, independent from the component logic and adaptable to the particular context. 
This model is achievable with a modular architecture, where capabilities can be plugged in and out dynamically, at runtime.

\subsection{The \cf architecture}

ComFlux is a proof-of-concept reconfigurable middleware to support the dynamic development of applications. It was originally designed to demonstrate the power of reconfiguration by way of `interactive experiences' in public spaces; though its functionality is more widely applicable.
The current implementation, examples and documentation are open source and available online at \url{www.comflux.net}. 

Since pervasive computing functionality emerges from the interactions between components, 
in order to work together, systems need to discover, understand, and communicate with each other. In distributed environments, functionality is driven by data and message exchange. 
Thus we built \cf as a messaging middleware, in which external control and communication mediation occurs with respect to data transfer.


\subsubsection{A messaging architecture}
\label{sec:messaging}
\cf is built as a framework for \emph{typed} message exchange between \emph{endpoints}. 
Applications define endpoints by specifying a \emph{communication type} and a \emph{message structure}. Both the sender and the receiver check the message against a \emph{message structure} (or \emph{schema}) that allows the specification of simple and complex types. 
Communication occurs between endpoints that are \emph{mapped} (\emph{connected}) and can have one of the following types: 
\begin{itemize}
\item \emph{source--sink}: unidirectional communication between a source (producer) and any number of sinks (consumers) 
\item \emph{request--response}: a client issues a request to a server which replies with one response; for mapping to be possible both request and response definitions must match
\item \emph{request--response+}: as above, but the server can issue one or more responses
\item \emph{streaming source--sink}: unidirectional communication source--sink of unstructured/streaming data. 
\end{itemize}

\cf can thus conform to various communication models: event driven (publish-subscribe), client-server, and streaming.
Endpoints' definitions contain one of these types along with a message structure for sources and sinks, and two structures for request and response endpoints. Streaming endpoints are unstructured and offer direct, end-to-end communication. 

To keep the discovery and endpoint messages generic, components maintain a (reconfigurable) \emph{manifest} that describes and identifies the component.
The manifest is the information shared externally to facilitate discovery and communication. It contains an updatable collection of information about itself, e.g., supported communication types, addresses, access methods, public keys or user-declared information, and its endpoints, e.g. their type, message structure and permissions. 

The  communication model in \cf is fully decentralised to suit heterogeneous, ad-hoc environments. While its functioning does not rely on a central or cloud-based system, \cf mediates the communication with such services in a uniform manner, by exposing an endpoint to the application layer. 
The manifest enables schema negotiation on mapping requests, and message type matching.
\subsubsection{The \cf middleware implementation}
Space does not permit a full description of the implementation. Full details can be found at \url{www.comflux.net}. An overview is given here.

\cf is written in C and offers developer interfaces in C and JavaScript. 
Messages are encapsulated in JSON and the type is specified with JSON schemas. 

\begin{figure}[ht]
\centering
  \includegraphics[width=.6\linewidth, trim={3.5cm 18cm 3.5cm .9cm},clip]{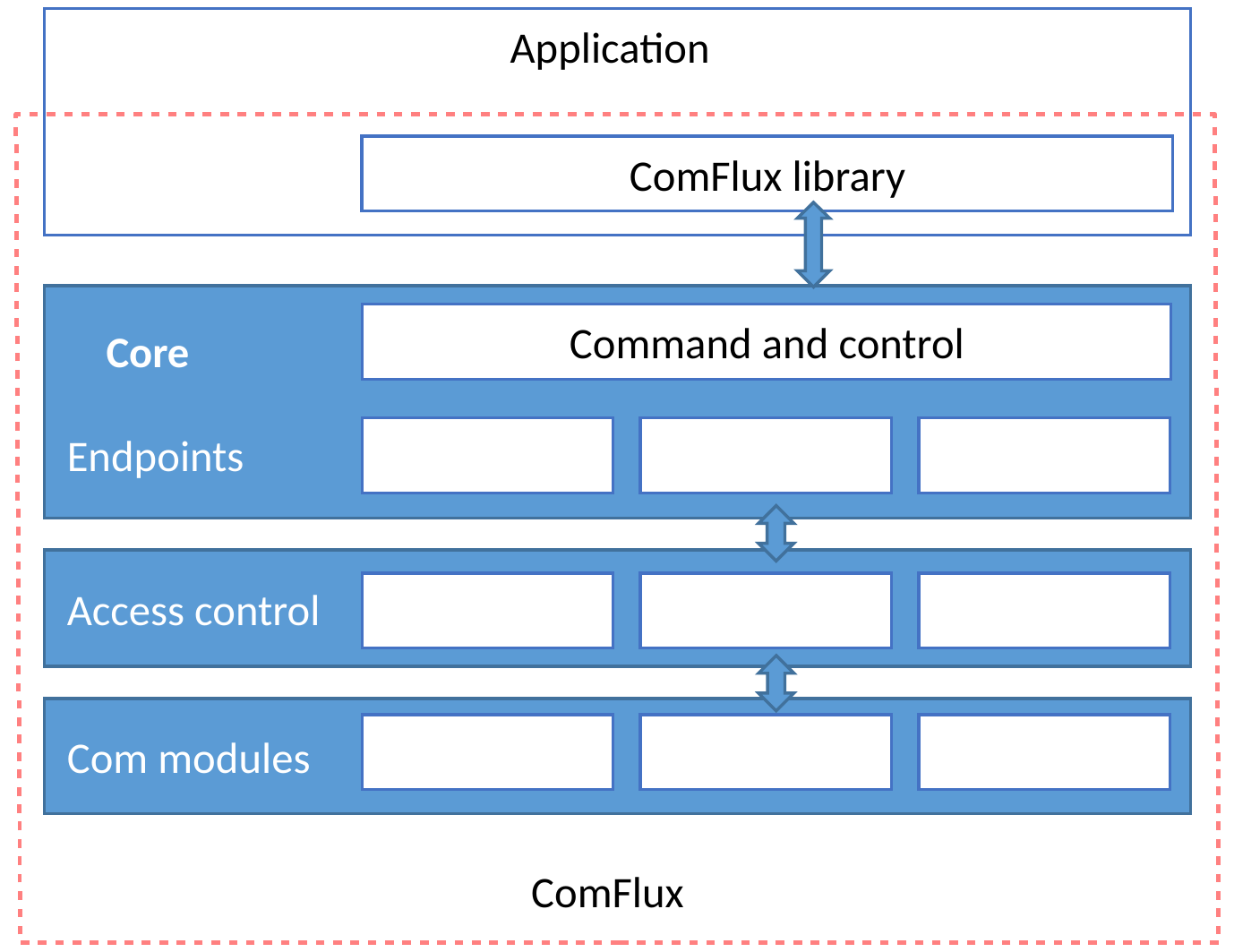}
  \caption{\small \cf architectural layers in a component}
\vspace{-2mm}
\label{fig:layers}
\end{figure}

 \cf  comprises the following layers, see Fig.~\ref{fig:layers}.
\paragraph{The core} is the backbone of the middleware and includes command and control functionality. It carries out data flow management and has the necessary mechanisms for  external reconfiguration. It decouples the management plane from the application logic by maintaining endpoints and orchestrating functionality.
\paragraph{Communication and access control modules} are functional units that plug into the core dynamically. They are reusable building blocks which are loaded and unloaded at runtime. Modules are built as dynamic libraries that implement an interface (a set of predefined functions) recognised by the core. \cf offers communication modules that carry the communication over a specific medium, and access control modules for authentication and endpoint access.
\paragraph{The \cf API} is available to the application programmer as a set of functions that provide internal endpoint and middleware configuration functionality. To use the middleware, the application imports the \cf library which automatically spawns the core process. The API provides the interfaces for sending\slash receiving data and configuring the core. 

Internally, the core has two logical layers: (1) the command and control layer and (2) the endpoint representation layer (see Fig.~\ref{fig:layers}). 
This exposes layer (1) in a uniform manner internally to the application, and externally, for other entities. 
The application manages the endpoints and accesses the middleware's functionality via the API. 
External control capabilities are enabled via \emph{control endpoints} which are defined automatically within the core, independent of the application logic.
%
\begin{figure}[bht]
  \centering
  \includegraphics[width=.9\linewidth, trim={1cm 10cm 5.9cm 2.3cm},clip]{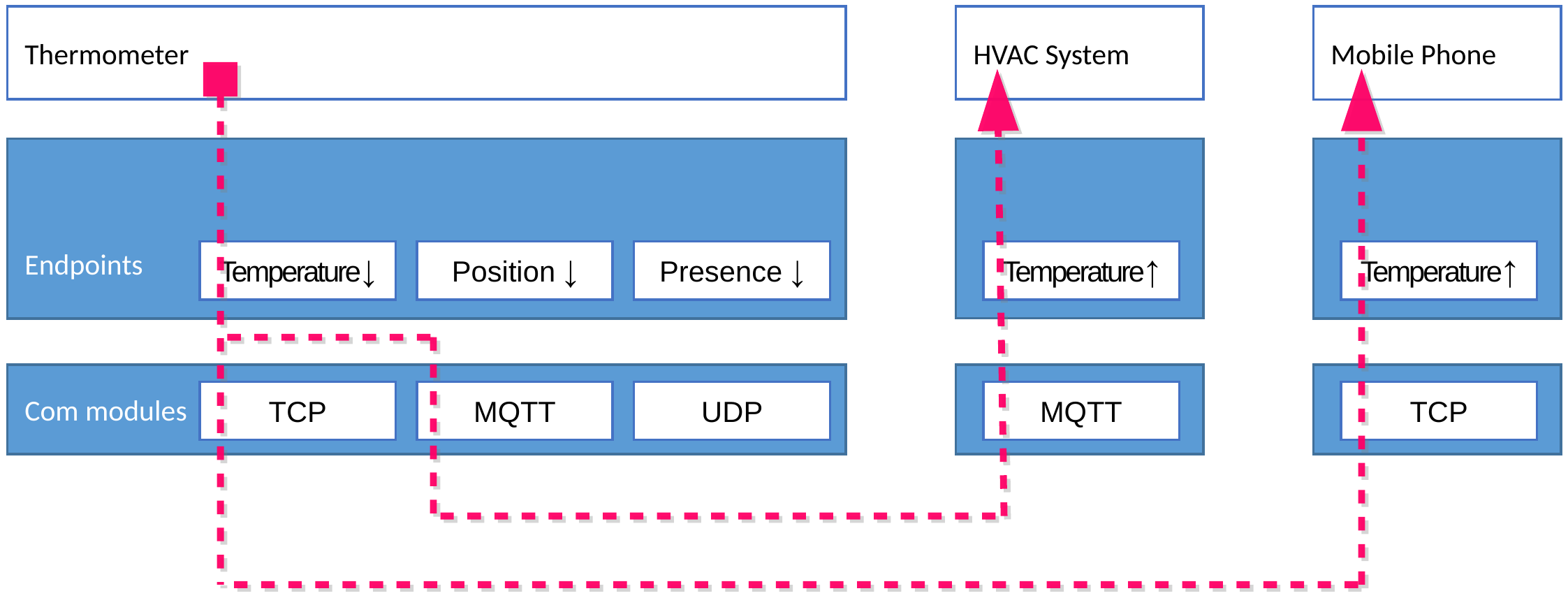}
  \caption{\small Data flow involving a thermometer component with a temperature data source ($\downarrow$) and two consumers ($\uparrow$), a HVAC system and a mobile phone. The HVAC system supports only MQTT while the mobile phone application uses TCP sockets.} 
\label{fig:flow}
\end{figure}
%
\subsubsection{Modules}
\label{sec:modules}
The power of \cf is in providing a mechanism to extend and adapt the functionality of an application at runtime. 
It achieves this with a module-based design, where functionality can be plugged into the middleware at runtime, internally, by the application and externally, through the control interface.
The core specifies interfaces for communication and access control.
Corresponding modules are built as dynamic libraries which are then loaded (linked) at runtime. 
This design choice allows them to be implemented as building blocks that can be reused with a new component and in a different application context.

\emph{Communication modules} may be either \emph{Bridge modules} that expect a middleware core/component on the other side of the communication channel (currently handling TCP, UDP and SSL); or \emph{Interfacing modules} that interface with other services (currently MQTT brokers and REST components).

\emph{Access control modules} implement an \emph{authentication} mechanism (currently username/password, SSL certificate or Kerberos) and maintain \emph{access control lists (ACL)} for each endpoint, describing the components that may interact with it.
\label{sec:access}

\subsubsection{Resource Discovery}
allows components to dynamically identify \emph{to whom} and \emph{how} to connect. 
To support \cf mappings we implemented \emph{Resource Discovery Components (RDCs)} which maintain a catalog of active, registered components.
A component can \emph{register} and periodically update its manifest with one or more RDCs so that it is discoverable by others.
The RDC provides a \emph{lookup} service, returning the addresses of components whose manifest and endpoint descriptions match the criteria in the query, e.g., a facilities management component may want to communicate with all light components in a building. If all the lights are registered with an RDC, the switch only needs to query that RDC for the lights of interest, allowing it to map (probe) the relevant devices in the building. Resource discovery is particularly important in interacting with mobile and dynamic components, which may `come and go' -- we explore this in Section~\ref{sec:demo}.
\subsubsection{Supporting external reconfiguration}
\label{sec:external}
is achieved by exposing \emph{control endpoints} that access the functionality in the control layer (see Fig.~\ref{fig:layers}). 
We make use of these capabilities with \emph{Swiss Knife}, our external command line tool. 
%
%
Control endpoints
behave and function like regular endpoints. Their messages are directly handled by the core functions that resolve API calls.
In this way both configuration and data functionality can be discoverable and negotiated at runtime. Loading/unloading modules, mediating connections, etc., can be executed by third parties at runtime.

\section{Demonstration of applications}
\label{sec:demo}
In this section we first describe how \cf could be used to realise a number of applications for smart cities, including the use of games to explore a city. 
We then show how externally controlled, dynamic reconfiguration leads to smooth integration and seamless interactions between people and the environment. As our work was in the context of infrastructure for enabling interactive public spaces (with some applications demonstrated in the Science Museum, London), the examples focus on experiential aspects, which we use to demonstrate functionality of wider, more general relevance.

\begin{figure*}[htbp]
\centering
\includegraphics[width=\linewidth, trim={1.4cm 9.8cm .8cm 2.3cm}, clip]{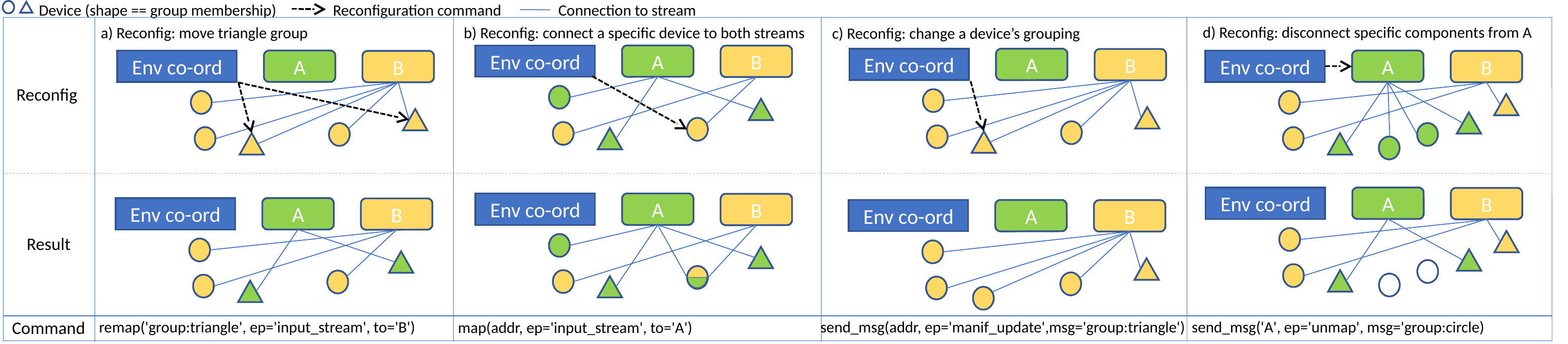}
\caption{\small A series of light show reconfigurations}
\label{fig:selectiona}
\vspace{-4mm}
\end{figure*}

\subsection{Smart cities}

We consider application scenarios within the cities of the future in which smaller scale systems  ---be they managed or ad-hoc architectures--- are composed, reused, and adapted to bring new functionality at a larger scale. 
We show how our reconfiguration framework can be used for the scenarios introduced in Section~\ref{sec:contextexamples}: seamless system composition that allows users to have services available wherever they go, and emerging pervasive gaming applications.

\paragraph{Seamless system composition} 
Current smart city and smart building applications are deployed as suites of technologies in discrete management domains, and do not currently work together. 
\cf would enable their connection to provide seamless integration within the whole smart city, where composition can lead to new ranges of public-sector, commercial and individually-customised applications.

Consider an application where at night, you exit public transport and walk away from other people towards home. Lights are activated, ``following you'' as you move towards your apartment building. As you arrive, the door opens and lights are again activated as you proceed to your own flat. The door opens as you arrive and enter your home, which is warm, due to prior notification of your arrival.

With \cf this scenario is enabled with external reconfiguration of access credentials (and possibly communication modules). On the streets and indoors, users will be equipped with their credentials according to their role.
The access in a workplace can be similarly enabled. In more sensitive areas, such as computing facilities, access may be allowed only between 9AM and 5PM, and only to administrators. 
With \cf to restrict access, a command will reconfigure the allowed credentials for the smart door lock at 5PM and will release it at 9AM. Exceptions can be issued by some authorised users or automatically, by emergency systems.

\paragraph{Enabling crowd applications: Pervasive gaming}
A way to discover a new city is through users' engagement with games.
With the pervasiveness of connected technologies, a world of \emph{mixed reality} emerges \cite{ Schouten2017, RicciPTC15, Spicer17, Montola2009}.  
Consider the following mixed reality game in which a player walks in a new city and discovers its history and hidden gems by interacting with its smart infrastructure. A series of lights guide the player at night to the city's most impressive streets, information is given about specific landmarks and visualisation tools present their history. 
The scenario can be framed as a treasure hunt or geocaching in which the player has to find objects or solve puzzles.
Local sensor data and real time visualisation of information can yield clues while the gameplay may indicate less crowded areas, event attendance, etc. 
Interaction with the infrastructure may be by shaking or tilting phones to control actuators, lights, doors.

Richer games could allow teams of tourist players to cooperate or compete in a virtual game on the streets of the modern city.
This scenario embodies the interaction between the environment and people at large scale.
A mixed reality game called \emph{Ghosts in the city} is presented in~\cite{RicciPTC15} and we envisage that \cf would enable this scenario by coordinating a city's lights' infrastructure, the user's presence and her augmented reality glasses or mobile phone.

Due to scale and heterogeneity, this scenario is not possible with static control, pre-programmed into the applications. 
Instead it needs to dynamically support new applications as they emerge.  
With \cf, the above scenarios can be enabled via reconfiguration operation to meet high-level goals. 

\subsection{Group coordination implementations} 
\label{sec:museumapp}
We consider two approaches to implementing group coordination: (1) where a component embedded in an environment manages the others in the space for some collaborative effect; (2) fully decentralised control and group coordination.

\emph{(a) Group-oriented light show:~} 
Inspired by the London 2012 Olympic Games ---where the main stadium was turned into a ``light canvas'', displaying patterns through a centrally-driven (stage-managed) orchestration of the lights on custom
devices held by the audience (attached to the seating)--- our first demonstrator application coordinates displays across people's mobile devices to create a landscape for visual effects.

Our demonstrator environment contained two components, each producing a unique stream of colour patterns. A device connected to one of these components would display the pattern it produced. To realise the desired `big picture' visual effects, 
another component, the \emph{environmental coordinator},  issues various reconfiguration commands to dynamically alter the devices connected to the streaming components.
To give greater control and better illustrate the infrastructure's management capabilities, the people/devices were split into two groups (defined by their metadata, set at runtime).

We illustrate some reconfigurations and the resulting effects in Fig.\ref{fig:selectiona}. Though a simple example,\footnote{Note that the effect of light patterns in the dark can be dramatic, and such an  approach might suit more complex and data-rich streams, e.g., sound (see below), video feeds, etc.} we include it to illustrate the ease and flexibility of reconfiguration commands in targeting and diverting groups of components, individual components, and those based on their current connections. 

\emph{(b) Silent Disco:}
A silent disco differs from a traditional club\slash disco since, rather than 
having a sound system (speakers), participants use headphones with an embedded channel selector. Two DJs each broadcast music on a separate channel, and each person with headphones uses the selector to decide the music they listen\slash dance to.  A coloured light on each headphone set indicates the channel one hears. The fun and spectacle is in the disconnect between those listening to different channels, differing in their dance moves and in `singing along'. 

Currently the experience is somewhat passive, in that one can only select the channel they are listening to. We consider a richer experience, by enabling technical interactions between participants in the space.
For example, a common action in silent discos is to have your friend(s) change to the channel you're listening to.
This can be rather disruptive, e.g., involving removing your own headphones to talk or even forcefully adjusting their selector.
Here we consider changing the channel of an individual or a specific group by means of an external reconfiguration command to their headphones\slash devices, to divert them to the desired stream.
Similarly, the DJs might decide at some stage to switch the whole crowd to a particular music stream, for instance, in building up to a finale. With the reconfiguration capabilities of \cf we can incorporate these features seamlessly in an engaging manner.

We include this example to highlight the capacity to manage participants and groups: participants might only be able to change the channels of their friends (rather than annoy strangers), while the DJs can interact with everyone. In our example, we are able to dynamically manage groups through component metadata and credentials (for each listening device), which is reflected in the RDC. This in turn allows a dynamic selection of peers within groups they belong, while limiting the visibility of those who do not belong and therefore by which interactions are barred. In Fig.~\ref{discorules} we present a series of reconfiguration rules for these interaction patterns.

\begin{figure}[t]
{\tiny 
\begin{tabular}{|@{~}>{\raggedright}p{1.2cm}@{~}|@{~}>{\raggedright}p{2.2cm}@{}|@{~}p{5.1cm}@{~}|}
\hline
 \bf{Function} &
 \bf{Reconfiguration command(s)} &
 \bf{Resulting actions (automatic via the middleware)} \\
\hline
Anne joins 
Bob's birthday 
group & 
manifest\_add(`group:birthday') credential\_add(`birthday') acl\_add(`music\_input', \\ ~~~~~`remap', `birthday') & 
Anne's device updates the metadata on groups.
This update is communicated to the RDC.
A credential for the group is added, to assist in managing group interactions.
Anne's access control list is also updated to authorise those in the `birthday' group (holding the credential) to alter (remap) the connections for her `music\_input' endpoint. \\
\hline
Bob changes Anne's channel to DJ Mikey (DJ1) &
remap(`person:Anne', \\ ~~~~~`music\_input', `DJ1') &
Bob's device queries the RDC to find the address of Anne's device, based on name and group membership.
Bob's device sends a command to remap Anne's `music\_input' endpoint to DJ1. Given Bob is also part of `birthday', Anne's device accepts and actions the command by disconnecting her endpoint from DJ2 and connecting it to the requisite endpoint (matched based on schema) on DJ1.\\
\hline
Anne changes the whole group to DJ Sandra (DJ2) &
remap(`group:birthday', \\ ~~~~~`music\_input', `DJ2') & 
Similar to the previous example, but here it is Anne's device that discovers and sends `music\_input' remapping commands to everyone in the `birthday' group.\\
\hline
As DJ Sandra is leaving, she diverts all her listeners to DJ Mikey (DJ1) &
divert(`*', `music\_output', \\ ~~~~~`DJ1') & 
Here DJ Sandra's music streaming component sends a remap command to every component connected to her `music\_output' endpoint (i.e., all her listeners). As a DJ, she holds credentials to alter the `music\_input' connections of all devices, regardless of their groupings. The result is each device disconnects from her, and connects to DJ Mikey.\\
\hline
\end{tabular}
}
\caption{\small Example silent disco reconfigurations}
\label{discorules}
\vspace{-5mm}
\end{figure}

This section has demonstrated implementations of centrally managed and distributed group interactions. We also discussed how new applications could be composed and existing applications adapt to new contexts.
Through these examples we have shown that with \cf, seamlessness and interaction do not rely on application design, instead they emerge from dynamic communication reconfiguration. The interaction is not limited to the set of initial devices and communication patterns.

\section{Performance evaluation}
\label{sec:eval}
We implemented a suite of demonstrators using \cf for \emph{Science Museum Lates}~\cite{lates} 
We now evaluate the performance of \cf in this scenario, 
to indicate the feasibility and practicality of
of embedding such a control infrastructure within a communications framework. 
We compare workloads between a \cf-enabled component and an application without the middleware, both using the same mode of communication.  
Our measurements address only the average communication overheads.

Key performance measurements for pervasive systems include resource usage and latency. 
In our use cases, throughput is important for real-time interaction.  Therefore our measurements will primarily focus on measuring latency of message transmission between applications. 
In addition, in targeted use cases, by using single threaded machines, latency directly reflects CPU consumption.
We also discuss memory usage, a generally limited resource in embedded and IoT systems.


\paragraph{Methodology}
Event driven architectures have become the de facto solution for reactive and interoperable IoT. Components listen to events published in a desired context and react accordingly. 
One of the most common scenarios for disseminating information is through indirect asynchronous communication in a publish/subscribe interaction model.  

It is often the case that a source regularly broadcasts data to the subscribers, for instance sensors regularly publishing their readings. 
As an illustration of this, we use the source-sink interaction model for our museum demonstrations, where components listen for changes in the environment.  
In a collaborative drawing application, participants use this interaction model on two endpoints: to broadcast their updates and to receive the group's updates on the canvas. 
We thus chose to evaluate \cf using a message transmission scenario between two applications, a source A and a sink B. 

The evaluation deployment runs as follows. A  source application A sends to the sink application B a message signalling the start of the transmission. Following the initial message, B starts the time counter and A begins sending $n$ messages at regular time intervals, every 10ms. After receiving $n$ messages, B records the total time for receiving the entire load (excluding the sleeping time). 
In the test components, the middleware is configured with the chosen communication module. The control application sends messages directly via the corresponding protocol interface. 
We employed BSD TCP sockets and the MQTT library~\cite{mqttrep, mqtt} with the Mosquitto broker~\cite{mosq}. 

In our test deployment we vary the load; the number of messages ranges over $n=500, 1k, 10k, 50k$ and the size of the message $s=.1KB, 3.5KB, 1MB$.
The small message corresponds to a GPS location and date-time value. The larger message contains a 3.5KB text message and the largest a 1MB message containing the encoding of a photograph.
%

Recall (Section \ref{sec:cf}) that our middleware includes type checking of messages and the enforcement of particular interaction paradigms. We therefore include message verification against a schema in the test.

\paragraph{Setup}
We deployed the case study applications on Amazon cloud virtual machines, so located in the same independent geographic area. 
Components were mapped across different machines in the same availability zone. 
We used virtual machines of type t2-micro in the same Amazon region, eu-west-1 Ireland but on different physical machines denoted eu-west-1b and eu-west-1c. 
Each virtual machine had 1GB of memory and a single virtual CPU implemented with 6 CPU credits on an Intel Xeon E5. 
The host OS was Ubuntu-trusty-14.04. Experiments were conducted between July 28th and August 5th, 2017. 
%
The applications create a JSON message. 
The small message is encapsulated as in Fig.~\ref{fig:msg}. 

\begin{figure}[ht]
\includegraphics[width=\linewidth, trim={2.5cm 20.2cm 2.5cm 1.5cm},clip]{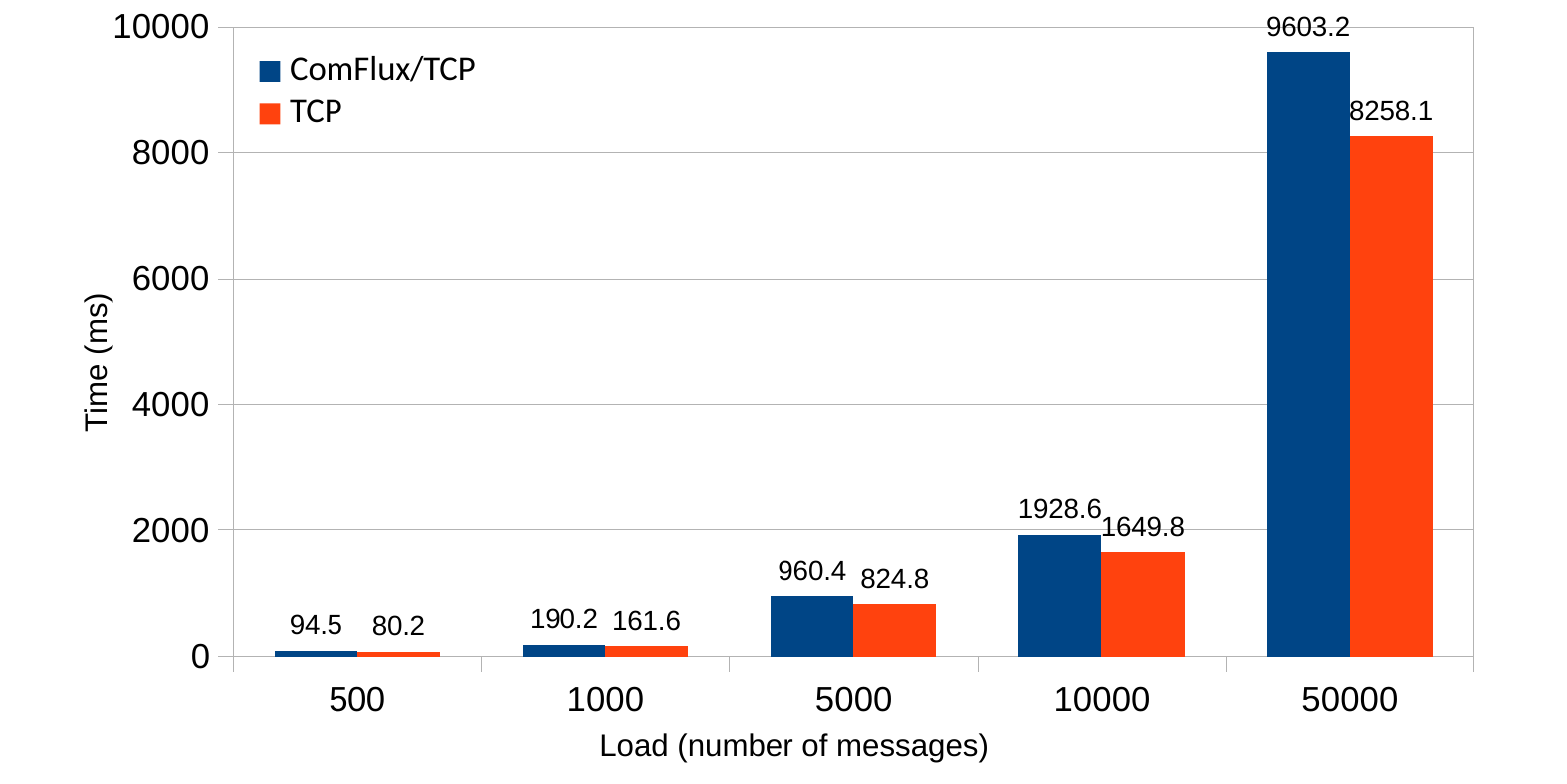}
\caption{\small Performance impact on TCP communication}
\label{fig:eval}
\end{figure}

\paragraph{Results}
The core occupies 176KB and the API libraries 120KB. As a messaging-passing infrastructure, the core does not generally store messages after they are pushed to the application\slash remote \cf instance, and the memory consumption remains constant through the execution. 
Fig.~\ref{fig:msg} shows a message containing the GPS and date-time information.
Each message is encapsulated into a JSON structure to facilitate message parsing by the receiver. The message set by the application is stored under the \texttt{msg\_JSON} attribute. Additionally the message contains an id (field \texttt{msg\_id} is an incremental integer) and a message type or \texttt{status} (in the example the message from a source to a sink has code 9).

\begin{figure}[htbp]
\begin{center}
{
\scriptsize  \begin{verbatim}
  { 
   "msg_id": "13",
   "status": 9, 
   "msg_JSON": {
     "position": "41.24'12.2\"N 2.10'26.5\"E", 
     "date": "2012-04-23T18:25:43.511Z" }
  }
\end{verbatim} }
\caption{\small GPS and date-time message encapsulated with \cf}
\label{fig:msg}
\vspace{-3mm}
\end{center}
\end{figure}

The performance results measure the overall time per message once the connection has been set up. On average, the middleware adds 16\% to performance overhead for a medium sized message. 
Fig.~\ref{fig:eval} illustrates the measurements for a 3.5 KB message using TCP communication. In our experiments the performance overhead for MQTT was also 16\%.
This value increases up to 25\% for smaller size messages and messages larger than 1MB. In the first case, the overhead is due to greater impact of the message size, while in the second, the main factor is the JSON parsing, which is more costly for larger messages.

\paragraph{A proof-of-concept implementation}
We built a fully-fledged middleware incorporating a rich range of functionality. 
The current implementation of \cf is meant to facilitate the deployment of broad functionality for interactive applications. 
\emph{The key contribution of this paper is to provide a practical demonstration of the potential for   
dynamic, external command and control capabilities. }
As such,  our focus was not on performance considerations, but on flexibility, reliability, and usability.
For instance, messages are type-checked at both the sender and the receiver. Additionally, the mapping process enforces negotiation of communication and access control mechanisms and compliance with a common interaction type. 
These measurements are therefore intended to demonstrate the feasibility of embedding a command and control infrastructure within a communications framework. 
Our implementation was developed to support interactive applications for museums, in a dynamic, ad-hoc, people-centric environment, and was designed accordingly.

\paragraph{Optimisations}
We plan several ways to optimise the performance. 
Firstly, transmission-related optimisations will be added to leverage per-message performance. Message type-checking can be simplified and even eliminated for a wide range of applications. While such a capability is the default for streaming interactions --communication is point-to point with non structured messages-- all other endpoint types require and check against a schema. 
For illustration, communication modules that interface with existing services (e.g., MQTT brokers or REST web services) do not need transmission to conform to a JSON message structure, as compliance can be handled at the module or application level. Furthermore, in performance-oriented application examples, the preference may be to avoid a message structure.
Secondly, we envisage \cf to run as a common system-wide middleware for all applications. Currently it is spawned by each application individually. 
Thirdly, \cf can leverage the performance of constrained embedded devices by running on a separate machine, in the local network. 

\section{Related Work}
\label{sec:rw}

Our approach complements existing work in context awareness, service composition and industry trends. 
Our vision is to integrate with research in pervasive computing on context aware, composable and adaptable systems, to facilitate the development and deployment of IoT applications.

\paragraph{Commercial solutions}
Consumer solutions attempt to integrate a large range of IoT devices into ``smart home'' platforms.
Apple HomeKit~\cite{homekit}, Amazon Alexa and Echo~\cite{alexa} and Google Home~\cite{home} are commercial solutions for managing multiple third-party devices under a single user interface, via a voice activated hub and a phone. While they integrate the most popular devices and services, they tend to focus on reinforcing their own technical ecosystem, often in a more centralised manager, rather than the system-wide control and coordination.

In terms of development platforms, many focus on exploiting event-driven architectures and involve the network stack, cloud services, and a suite of APIs for web or smartphone. They require an account registration and sometimes a periodic fee. 
There are commercial PaaS solutions (Apple HomeKit~\cite{homekit}, ThingWorx~\cite{thing}, Kaa~\cite{kaa}, Xively~\cite{xively}, etc.), 
development frameworks (IoTivity/AllJoin~\cite{iotivity}) and embedded OSs (e.g., Contiki~\cite{contiki}, mbed OS~\cite{mbed}). 
Typically, the frameworks rely on the cloud for device management and do not always integrate easily with existing solutions (vendor `lock-in', reinforces a technical ecosystem). But even when they allow decentralised deployment, they still do not support external control --- rather, the configuration and preferences tend to be specified by developers, embedded in application code.
As we have argued, the wider pervasive vision requires more, including the flexibility to externally and at run-time, adapt system interactions across applications. 

IFTTT~\cite{ifttt}, is a service allowing the creation of a `recipes' (a series of conditional statements), that trigger specific API functions of various components and services in response to particular events. 
Though providing a useful and accessible way forward in enabling the coordination and interaction of disparate components, it functions as `API-glue'; the possible functionality is limited to that exposed by the specific APIs of the components, and there is no consistent management framework in any reconfiguration is limited to what each component developer decides to explicitly include in their API.
Moreover, the often closed and opaque tendencies of many commercial offerings are harder to incorporate into research.



\paragraph{IoT Management: adaptation and composition} 
Management of components in heterogeneous environments remains a research challenge \cite{percomvision, CaceresF12, Alur16}. 
The vision is that functionality should be encapsulated in modules, then composed and reused in different circumstances \cite{MakrisSS13, ZhouGPRYS11, BealPV15}, but has yet to be fully realised. 

Aggregate programming \cite{BealPV15} is a paradigm for IoT development to facilitate design, creation, and maintenance of IoT systems. A proposed model for the development of complex components is that basic, distributed components ---such as sensing, communication, localisation--- are encapsulated and composed with one another into building blocks that can be adapted and reused.
\cf relies on a similar overarching principle: system composition of heterogeneous building blocks. Moreover in \cf, composition can be performed ad-hoc, by external control, in response to triggers or commands from the environment.

SBUS is a stream-management infrastructure that has been used as a testbed for city-wide transport monitoring, healthcare and lifestyle management \cite{SinghB14}. 
Our work on \cf redefines SBUS, as is necessary for the broader vision of pervasive computing. Specifically, \cf introduces a modular architecture, which not only extends the reconfiguration capabilities and increases flexibility, but also assists the longevity of deployments by enabling reconfiguration when\slash where necessary.

\paragraph{Context awareness}
has been understood as the capacity of a component to acquire information from and reason about the environment, then adapt its behaviour accordingly. Most of the proposed solutions collect data from a limited number of physical (hardware) and virtual (software) sources \cite{PereraZCG14} and focus on retrieval of data, with data management in the middleware layer.  
A great challenge in pervasive computing has been the management of components in the environment~\cite{HarterHSWW02, CaceresF12, MakrisSS13, PereraZCG14}. 
The early work in \cite{HarterHSWW02} describes a small-scale indoor application for sensor-driven context-aware computing that enables applications to follow mobile users as they move around a building.

Targeting the development of adaptive systems, \cite{YauK04} proposes RCSM, a middleware and interface for specifying context-aware interfaces on the same device.
It is adaptive in the sense that it changes its management and resource discovery to those of the local applications, but it does not communicate externally.

For an Android smart-phone, the work in \cite{ElmalakiWS15} proposes decoupling application logic from complex adaptation decisions.
CAreDroid uses a set of context-sensitive functions that offer runtime support for multiple, concurrent, context-aware applications accessing sensors.   
In \cite{AgirrePAEM16} the authors propose a framework that supports reconfiguration for both internal (application triggered, API calls) and external (application unaware, contextual) events.
For both types of trigger, the reconfiguration needs to recognise external triggers and program them into the middleware logic. 
In our approach, the reconfiguration is direct and data-driven, by connecting and interacting with the component.
 
The authors in \cite{LopesGSDSCBPYG13} address similar problems of pervasive environments: heterogeneity, wide distribution, dynamism, and mobility.
Their work presents an architectural model for context awareness that supports contextual data acquisition, control over functionality within the environment, and processing of contextual information. However, this approach relies on  central components (servers) to maintain and process contextual information, including handling actuation commands. Furthermore, the design does not allow runtime management of middleware modules. 

In \cite{ZhouGPRYS11} the authors identify the following context-based changes: peer coordination, process-service adaptation and utility-service adaptation. In the proposed system, Context-Aware Pervasive Service Composition (CAPSC), a composition adaptor receives commands resulting from context reasoning, then chooses and performs the service composition. Composition and adaptation are performed indirectly and triggered internally as a result of reasoning about the environment.  

In the above examples, applications react to context triggers in a programatic way. In contrast, \cf enables contextual adaptation, by allowing---given the requisite permissions---direct control, management and mediation over components through the communications infrastructure (messaging), rather than in code. This allows flexible, ad-hoc and dynamic adaptations, capable of supporting customisation, personalisation, and evolution over time with minimal intervention overheads.

\paragraph{Smart Cities and pervasive gaming} 
There is interest among researchers and creative industries' professionals in the development of games~\cite{Cowling16, RicciPTC15, Spicer17}, which we consider a direct application of \cf. 
In a smart city scenario, \cite{Spicer17} demarcates what future games might be like: interacting seamlessly with the environment, using public infrastructure in new ways and engaging citizens. 
Our contribution is a paradigm shift from API-based applications, where our external command and control will help development and deployment of interactive connected worlds, offering far richer experiences by allowing the integration of existing (\cf dedicated) infrastructure. 

\paragraph{Privacy and security} 
Pervasive computing ---with an emphasis on consumer IoT--- is envisaged to become open, dynamic, and involve many actors with different motivations and capabilities. Security and data privacy receive considerable attention~\cite{LuoFPKGLSK16, SinghBE14}.
Access control regarding devices' capabilities, and the exchange of data is fundamental, and we believe that our approach an open, modular access control regime, enabling permissions to be dynamically managed will be an important mechanism for assisting in  managing the security concerns of a highly dynamic environment. 

%
%
%
A related concern is the legal and compliance concerns of the IoT. This was considered in \cite{SinghPBPDE16}, which suggested that a legally-compliant IoT can be enabled through policy enforcement (control and audit) over data exchange. \cf provides some of the capabilities necessary for realising this, which we will explore in future work.

Finally, though pervasive computing is an active area of research, not always is there code available for the community to take forward. An explicit aim of our work is to encourage use, experimentation and extension by the research community, by making the \cf platform (source and documentation) openly available (\url{www.comflux.net}). 

\section{Concluding remarks}
\label{sec:concl}

The vision for pervasive computing involves enabling 
contextual functionality adapted to suit the individuals and integrated seamlessly with a dynamic environment.
We have argued that key to realising  this vision 
is a modular approach in which components are composed and coordinated across management ``silos'' by enabling runtime external management of component connections and capabilities.

We introduced in this paper \cf, a modular infrastructure that demonstrated the feasibility of an external management regime. 
We also illustrated the practical considerations and proposed an implementation which demonstrates the feasibility of our approach with acceptable performance overhead.

Our approach is intended as a tangible step forward in enabling a range of new functional possibilities as presented in Section \ref{sec:context}. 
We have already shown that a general implementation aimed to support interactive exhibits in cultural and event spaces, see Section~\ref{sec:demo}. 
By making the \cf source available, and describing its conceptual underpinnings, our aim is to provide the community with a practical basis for an external command-and-control regime, to move towards the broader vision of pervasive computing.

\bibliographystyle{IEEEtran}
\clearpage 
\begin{spacing}{1.1}
\Large
\bibliography{bibliography}
\end{spacing}

\end{document}